\documentclass[11pt]{article}
\usepackage{latexsym}  
\usepackage{amssymb}
\usepackage{graphicx}
\usepackage{amsmath}

\begin{document}

\begin{center}
{\Large\bf Neutrino Mass Hierarchies }\\

{\Large\bf in a Mass Matrix Form Versus its Inverse Form}

\vspace{3mm}
{\bf Yoshio Koide}

{\it IHERP, Osaka University, 1-16 Machikaneyama, 
Toyonaka, Osaka 560-0043, Japan} \\
{\it E-mail address: koide@het.phys.sci.osaka-u.ac.jp}

\date{\today}
\end{center}

\vspace{3mm}
\begin{abstract}
A neutrino mass matrix model $M_\nu$ with $M_\nu^T =M_\nu$
and a model with its inverse matrix form 
$\widetilde{M}_\nu = m_0^2 (M_\nu^*)^{-1}$ can be diagonalized
by the same mixing matrix $U_\nu$. 
It is investigated whether a scenario which provides a matrix 
model $M_\nu$ with normal mass hierarchy can also give a model 
with an inverted mass hierarchy by considering a model with 
an inverse form of $M_\nu$.  
\end{abstract}

\vspace{3mm}

 
It is very interesting problem in the current neutrino physics 
whether neutrino mass hierarchy is normal or inverted.
If the observed neutrinos are of Majorana type, 
models with an inverted hierarchy will be testable by observing 
the effective neutrino mass $\langle m_{ee}\rangle$ in the near 
future neutrinoless double beta decay experiments 
because we expect $\langle m_{ee}\rangle \sim
\sqrt{\Delta m^2_{atm} } \simeq 0.05$ eV.
On the other hand, if we cannot observe a non-vanishing value of 
$\langle m_{ee}\rangle$ up to $\sim 0.01$ eV, as far as the 
hierarchy is concerned, its test will be hopeless 
in the near future experiments. 
For the prediction whether the neutrino mass hierarchy
is normal or inverted, current models are half and half.
However, in this paper, we notice that, in a model 
with a mass matrix form $M_\nu$ ($M_\nu^T =M_\nu$) which can 
provide reasonable masses $m_{\nu i}$ ($i=1,2,3$) and mixing 
$U_\nu$, a mass matrix with an inverse form  
$\widetilde{M}_\nu = m_0^2 (M_\nu^*)^{-1}$ can 
give the same mixing $U_\nu$ as in the original model $M_\nu$.
This means a possibility that a model with a normal 
mass hierarchy can also give an inverted hierarchy.
Then, most current models will be able to predict
an inverted hierarchy, i.e. most models have a possibility
 $\langle m_{ee}\rangle \sim 0.05$ eV.
This will encourage experimental physicists who have a plan
to measure $\langle m_{ee}\rangle$ with the order of 0.05 eV.

Such a scenario with ``dual" hierarchy is the following case:
Let us consider an effective neutrino mass matrix $M_\nu$ 
on a flavor basis on which the charged lepton mass matrix $M_e$ 
is diagonal (or nearly diagonal).
The mass matrix form $M_\nu$ is invariant under a flavor
basis transformation $T$, i.e.
$$
T^T M_\nu T = M_\nu .
\eqno(1)
$$
We may consider such operators $T$ more than two.
In other words, we consider that the mass matrix $M_\nu$ has
been derived from the constraints (1). 
Since we know a no-go theory \cite{no-go} for flavor symmetries,
we consider that the mass matrix $M_\nu$ is an effective type,
e.g. the form $M_\nu$ is described in terms of Higgs scalars
more than two, or it is of a Froggatt-Nielsen type 
\cite{Froggatt}.
(Therefore, the matrices $T$ can operate only to neutrino
sector independently of the charged lepton sector.)
We also define a neutrino mass matrix $\widetilde{M}_\nu$ as
$$
\widetilde{M}_\nu = m_0^2 (M_\nu^*)^{-1} + m_0\xi_0 {\bf 1}.
\eqno(2)
$$
From the inverse of Eq.(1), we obtain 
$$
T^T (M_\nu^*)^{-1} T = (M_\nu^*)^{-1} ,
\eqno(3)
$$
so that
$$
T^T \widetilde{M}_\nu T = m_0^2 (M_\nu^*)^{-1} + m_0 \xi_0 T^T T.
\eqno(4)
$$
Therefore, for a case that the matrices $T$ are orthogonal (e.g.
an exchange operator between $\nu_\mu$ and $\nu_\tau$ in the 
2-3 symmetry \cite{23sym}, and so on),  we obtain 
$$
T^T \widetilde{M}_\nu T = \widetilde{M}_\nu, 
\eqno(5)
$$
i.e. the mass matrix form $\widetilde{M}_\nu$ is also allowed under 
the same symmetry.
(Note that for operators $T$ with $T^T T \neq {\bf 1}$ the matrix
$\widetilde{M}_\nu $ is not invariant under the symmetry. 
For such a general case with $T^T T \neq {\bf 1}$, 
 the invariance (5) holds only when $\xi_0=0$. 
On such a case, we will comment later.)

On the other hand, the neutrino mass matrix $M_\nu$ is diagonalized as
$$
U_\nu^T M_\nu U_\nu = D_\nu \equiv {\rm diag}(m_1,m_2,m_3).
\eqno(6)
$$
Therefore, the matrix $\widetilde{M}_\nu $ is also diagonalized as
$$
U_\nu^T \widetilde{M}_\nu U_\nu = m_0^2 (D_\nu^*)^{-1} + 
m_0 \xi_0 U_\nu^T U_\nu = {\rm diag}(\widetilde{m}_1,
\widetilde{m}_2, \widetilde{m}_3), 
\eqno(7)
$$
where we have assumed that $M_\nu$ is real matrix, so that 
$U_\nu^T U_\nu= {\bf 1}$.
Therefore, we can obtain the same neutrino mixing matrix $U_\nu$
for $\widetilde{M}_\nu$, too.
The mass eigenvalues $\widetilde{m}_i$ of $\widetilde{M}_\nu$ 
are given by
$$
\widetilde{m}_i = m_0 \left( \frac{m_0}{m_i} +\xi_0 \right) .
\eqno(8)
$$
When we define the ratio $R$
$$
R \equiv \frac{m_2^2-m_1^2}{m_3^2-m_2^2} ,
\eqno(9)
$$
correspondingly to the observed ratio 
$R_{obs} = \Delta m^2_{solar}/\Delta m^2_{atm}$, 
we obtain
$$
\widetilde{R} \equiv \frac{\widetilde{m}_2^2-\widetilde{m}_1^2}{
\widetilde{m}_3^2-\widetilde{m}_2^2 }
= \left( \frac{m_3}{m_1}\right)^2
\frac{1+2\xi_0 m_2 m_1/(m_2+m_1)m_0 }{1+2\xi_0 m_3 m_2/(m_3+m_2)m_0} R .
\eqno(10)
$$
Therefore, we can, in general, fit the value of $\widetilde{R}$ 
to the observed value by adjusting the free parameter 
$\xi_0/m_0$ suitably.
In fact, we can obtain $\widetilde{R}=\mp R$ by choosing 
the parameter $\xi_0/m_0$ as
$$
\frac{\xi_0}{m_0} = 
\frac{(m_3^2\pm m_1^2)(m_3+m_2)(m_2+m_1)}{
2 m_3 m_2 m_1 [m_3 (m_3+m_2)\pm m_1(m_2+m_1)] } .
\eqno(11)
$$

We have an interest in whether a model $\widetilde{M}_\nu$
gives an inverted mass hierarchy (IH) or not when the original
model $M_\nu$ gives a normal mass hierarchy (NH).
For convenience, we define that $m_2^2>m_1^2$ (so that the
model gives $\tan^2\theta_{solar} \sim 0.5$), so that a case 
with NH gives $R>0$.
Since the mixing matrix $U_\nu$ is identical both for
$M_\nu$ and $\widetilde{M}_\nu$, the masses $\widetilde{m}_i$ 
have also to satisfy the relation 
$\widetilde{m}_2^2 >\widetilde{m}_1^2 $ because $\widetilde{M}_\nu$
will also give $\tan^2\theta_{12} \sim 0.5$.
Therefore, $\widetilde{R}$ takes $\widetilde{R}>0$ or
$\widetilde{R}<0$ according as NH or IH in $\widetilde{M}_\nu$
as well as those in $M_\nu$.
In order to see whether the model $\widetilde{M}_\nu$
gives NH or IH, we calculate
$\widetilde{x} \equiv \widetilde{m}_2/\widetilde{m}_3$
correspondingly to $x\equiv m_2/m_3$.
If we see $\widetilde{x}^2 \ll 1$ ($\widetilde{x}^2 \gg 1$) 
for $x^2 \ll 1$, the case describes a model with NH (IH).
We can describe the value $\widetilde{x}$ as a function of $x$
by using the constraint (11):
$$
\widetilde{x}= - \frac{ \frac{m_2-m_1}{m_2+m_1} + \frac{m_2-m_1}{m_3}
+\frac{m_2 m_1}{m_3^2} }{ 1 +
\frac{m_2^2+m_1^2}{(m_2+m_1)m_3} -\frac{m_2 m_1}{m_3^2} } ,
\eqno(12)
$$
for $\widetilde{R}/R=+1$, and
$$
\widetilde{x}= - \frac{ \frac{m_2-m_1}{m_2+m_1}
\left(1+\frac{m_2}{m_3}\right) + \frac{m_2}{m_3^2}\left( 1-
\frac{m_2}{m_3} \right) }{ 1 +
\frac{(m_2-m_1) m_2}{(m_2+m_1)m_3} + 
\frac{(m_2-m_1) (2m_2+m_1) m_1}{(m_2+m_1)m_3^2}
-\frac{m_2 m_1^2}{m_3^3} } ,
\eqno(13)
$$
for $\widetilde{R}/R=+1$.
The value of $\widetilde{x}$ is highly sensitive to the
explicit value of $x$, so that we illustrate 
the behavior of $\widetilde{x}$ versus $x$ in Fig.1.
Note that since $m_1^2 = (1+R) m_2^2-R m_3^3$ from the
definition (9), we obtain a constraint
$$
x^2 > \frac{R}{1+R} ,
\eqno(14)
$$
for $R>0$ (NH), e.g. $|x| > 0.165$ for the observed value
$R=0.028$ \cite{MINOS,KamLAND}.
(For a case $R<0$ (IH), the relation (14) is always satisfied, 
so that such a constraint does not appear.) 
In Fig.1 (a), (b) and (c), we have shown only cases in which 
the original model $M_\nu$ gives NH ($x^2 <1$ and $R>0$) and 
the inverse matrix model $\widetilde{M}_\nu$ gives  NH 
($\widetilde{x}^2 <1$ and $\widetilde{R}/R=+1$) or IH 
($\widetilde{x}^2 >1$ and $\widetilde{R}/R=-1$) under the
constraints $\Delta m^2_{21}>0$ and $\Delta \widetilde{m}^2_{21}>0$.
As seen in Fig.1 (b), we find that we can always choose 
the value of $\widetilde{x}=\widetilde{m}_2/\widetilde{m}_3$
which can give IH with the same value $|\widetilde{R}|$ as $R$
by adding a suitable shift-term $m_0 \xi_0 {\bf 1}$ to the inverse 
matrix $m_0^2 (M_\nu^*)^{-1}$. 
Since the case (c) in Fig.1 gives $1/\widetilde{x} > 0.3$
($1/\widetilde{x} \sim 0.3$ at $x\sim 0.2$), the case 
may be called as a case of a nearly degenerated hierarchy (DH)
rather than a case of IH.

In this paper, we have interested in a case of NH$\rightarrow$IH.
An inverse case, IH$\rightarrow$NH, can be guessed from the case of
NH$\rightarrow$IH [Fig.1 (b) and (c)].
For reference, we illustrate the remaining case, 
IH$\rightarrow$IH, in Fig.1 (d).
As seen in Fig.1, the cases NH$\rightarrow$IH and
IH$\rightarrow$NH take place in the case with
$m_1/m_2 <0$, and the cases NH$\rightarrow$NH and
IH$\rightarrow$IH take place in the case with
$m_1/m_2 >0$.

So far, we have investigated a model under the constraints 
$T^T T ={\bf 1}$ and $U_\nu^T U_\nu ={\bf 1}$.
However, the existence of such constraints narrows   
applicable cases of our statement to models.
Finally, we would like to comment on a model with $\xi_0=0$.
Then, a mass matrix $M_\nu$ can always provide the inverse matrix 
model $\widetilde{M}_\nu \equiv m_0^2 (M_\nu^*)^{-1}$ which 
satisfies the same flavor symmetry 
$T^T  \widetilde{M}_\nu T =\widetilde{M}_\nu$ and which has 
the same mixing matrix $U_\nu$ independently of whether 
$T^T T \neq {\bf 1}$ and $U_\nu^T U_\nu \neq {\bf 1}$ or not.
The problem is whether such an inverse matrix $\widetilde{M}_\nu$
can also give a reasonable mass ratio $\widetilde{R}$ or not.
As seen in Eq.(10), such a case gives
$$
\widetilde{R} = \left( \frac{m_3}{m_1} \right)^2 R ,
\eqno(15)
$$
so that it seems to be ruled out because of the factor
$(m_3/m_1)^2$ in Eq.(15).
It is true as far as we want a model with an exact 
relation $|\widetilde{R}|=|R|$.
However, the value of $\Delta m_{21}^2 \equiv m_2^2-m_1^2$
is highly sensitive to the renormalization group equation
(RGE) effects and some unknown loop corrections in a 
beyond-standard model.  
[The magnitudes of the effects are dependent on 
the model (the so-called $\tan\beta$ in SUSY model, and so on), 
so that the effects are not always conclusive.]
For example, for a model of $M_\nu$ in which a predicted
value of $\Delta m^2_{21}$ at a higher energy scale is too 
small value compared with 
the observed value of $\Delta m^2_{solar}$ and a predicted
value of $(m_3/m_1)^2$ is not so large, it can be possible 
that the inverse matrix model $\widetilde{M}_\nu$ gives a
reasonable value of $\widetilde{R}$.
Therefore, we may expect that some of models $M_\nu$ can provide
a reasonable masses and mixing for the inverse matrix model 
$\widetilde{M}_\nu$, too.

In conclusion, we have investigated whether a scenario which 
provides a matrix model $M_\nu$ with normal mass hierarchy can 
also give a model with an inverted mass hierarchy by considering 
a model with an inverse form of $M_\nu$. 
If the model $M_\nu$ is derived under a flavor symmetry 
$T^T M_\nu T= M_\nu$ with $T^T T ={\bf 1}$, the inverse 
matrix model $\widetilde{M}_\nu$ which is defined by (2) 
can also satisfy the same symmetry as 
$T^T \widetilde{M}_\nu T= \widetilde{M}_\nu$, 
so that the scenario can give both model $M_\nu$ and
$\widetilde{M}_\nu$.
Then, we have find that if a model $M_\nu$ with normal 
hierarchy gives $m_1/m_2<0$, the inverse matrix model 
$\widetilde{M}_\nu$ can give a reasonable value of 
$\widetilde{R}$ with an inverted hierarchy.
Even for a case without the constraints $T^T T={\bf 1}$
and $U_\nu^T U_\nu ={\bf 1}$ (so that $\xi_0=0$ in 
Eq.(2)), we still have a possibility that the inverse
matrix model $\widetilde{M}_\nu$ gives a reasonable
value of $\widetilde{R}$ although we cannot give
the exact relation $|\widetilde{R}|=|R|$.
We may expect the observation of $\langle m_{ee}\rangle
\sim 0.05$ eV in the near future neutrinoless double
beta decay experiments by considering their inverse
matrix models for a considerable number of current models.

\vspace{6mm}

\centerline{\large\bf Acknowledgment} 
This work is supported by the Grant-in-Aid for
Scientific Research, Ministry of Education, Science and 
Culture, Japan (No.18540284).



\newpage
\begin{figure}[h]
\begin{center}

\begin{picture}(180,150)
\put(0,0)
{\scalebox{0.55}{\includegraphics{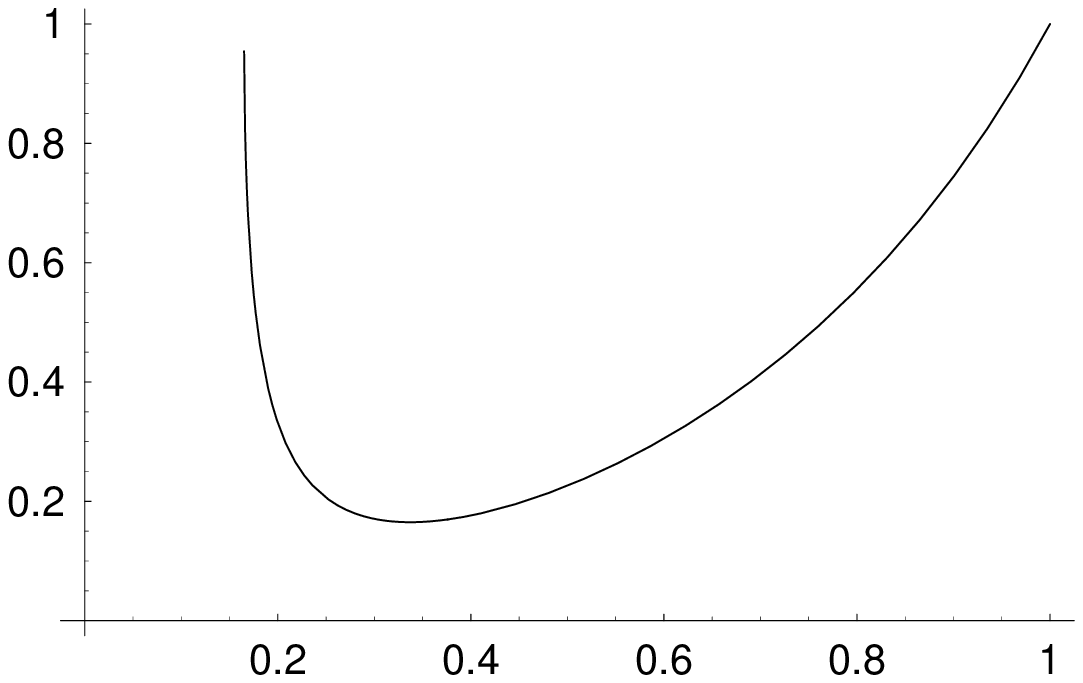}} }

\put(15,115){\large $\widetilde{x}$}
\put(180,8){\large $x$}
\put(50,-20){(a) Case NH$\rightarrow$NH}
\end{picture}
\hspace{10mm}
\begin{picture}(180,150)
\put(0,0)
{\scalebox{0.55}{\includegraphics{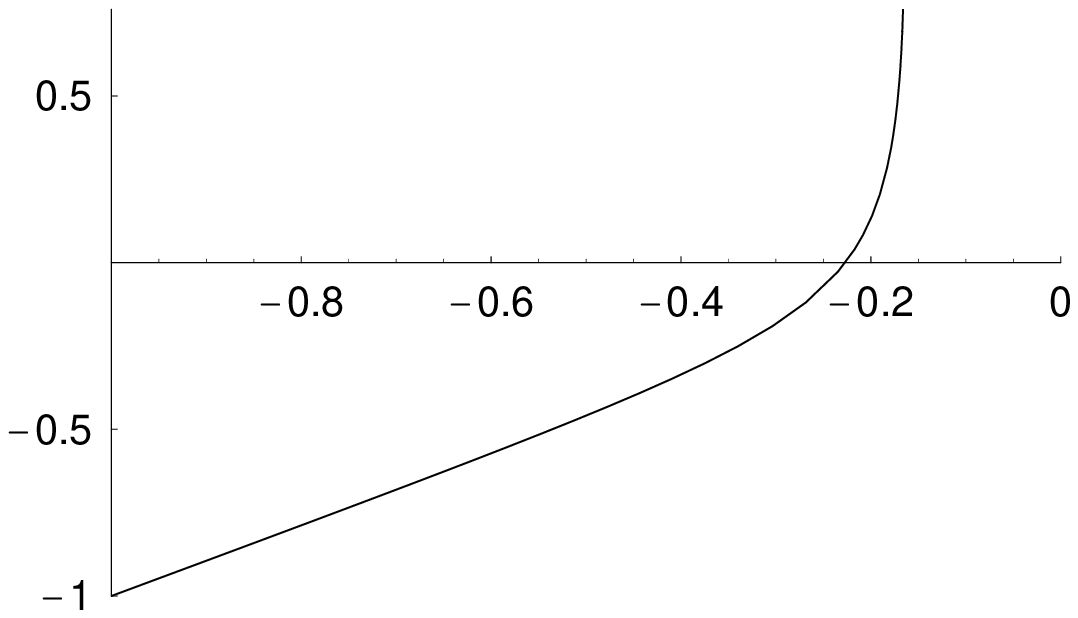}} }

\put(10,110){\large $1/\widetilde{x}$}
\put(180,60){\large $x$}
\put(50,-20){(b)  Case NH$\rightarrow$IH }
\end{picture}

\hspace{30mm}

\begin{picture}(180,150)
\put(0,0)
{\scalebox{0.55}{\includegraphics{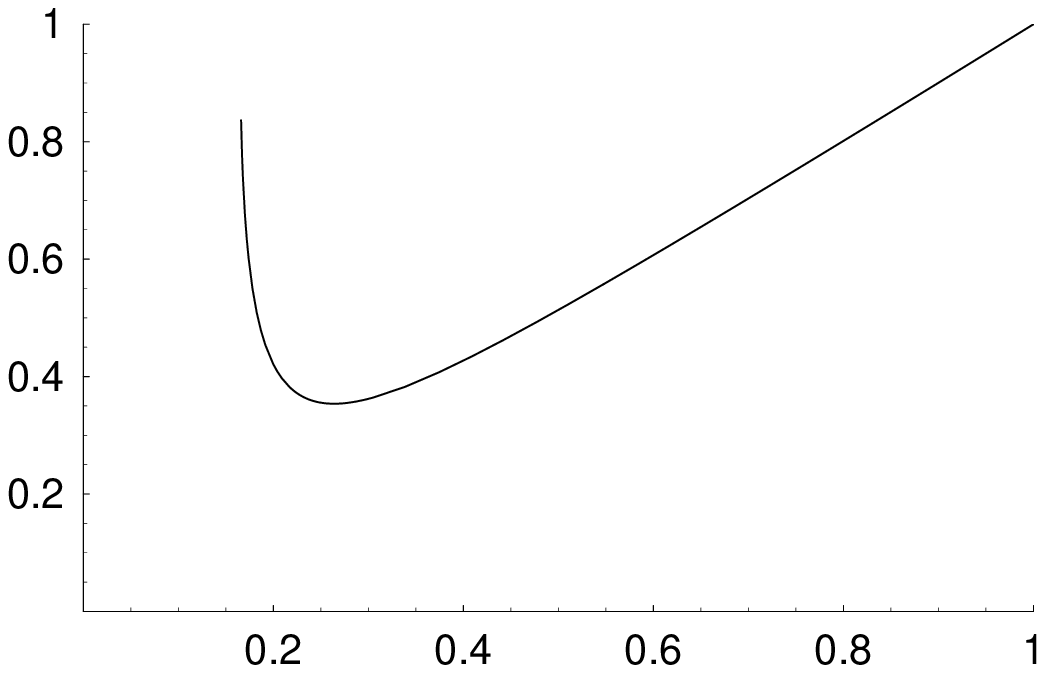}} }

\put(10,115){\large $1/\widetilde{x}$}
\put(180,8){\large $x$}
\put(50,-20){(c) Case NH$\rightarrow$IH}
\end{picture}
\hspace{10mm}
\begin{picture}(180,150)
\put(0,0)
{\scalebox{0.55}{\includegraphics{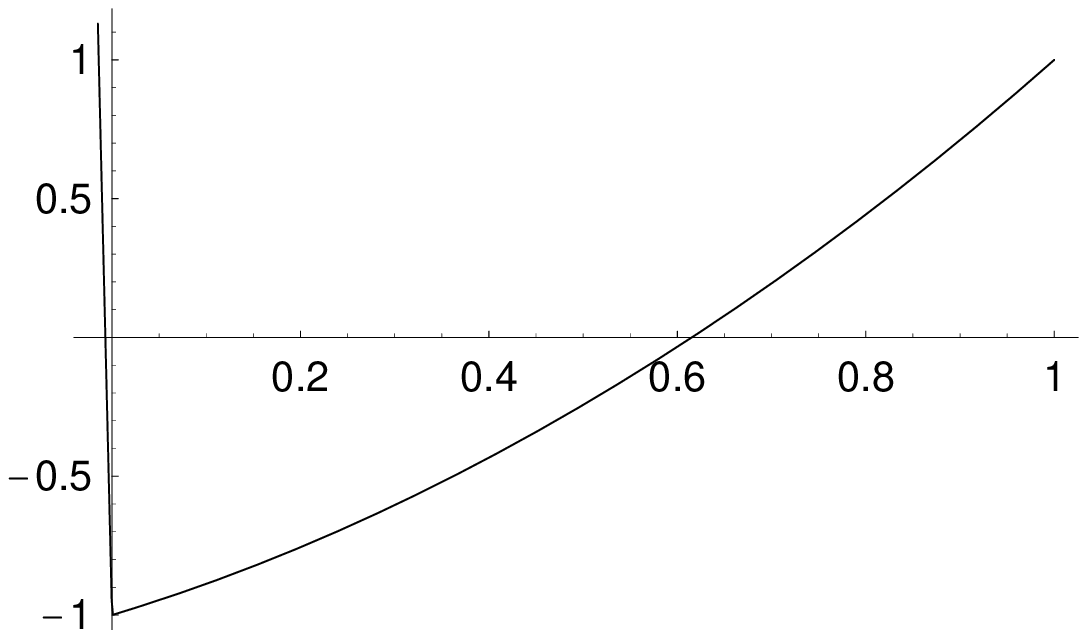}} }

\put(10,110){\large $1/\widetilde{x}$}
\put(180,50){\large $1/x$}
\put(50,-20){(d)  Case IH$\rightarrow$IH }
\end{picture}

\hspace{30mm}

\begin{quotation}
{\bf Fig.~1}  Behaviors of $\widetilde{x}=\widetilde{m}_2/\widetilde{m}_3$
versus $x=m_2/m_3$ under the requirement of 
$|\widetilde{R}|=|R|$: 
Input value $R=+0.028$ has been used as a NH case in
$M_\nu$. (a) a case of NH$\rightarrow$NH in $m_1/m_3>0$ and $m_2/m_3>0$; 
(b) a case of NH$\rightarrow$IH in $m_1/m_3>0$ and $m_2/m_3 <0$; 
(c) a case of NH$\rightarrow$IH in $m_1/m_3 < 0$ and $m_2/m_3>0$;
(d) a case of IH$\rightarrow$IH in $m_1/m_3 > 0$ and $m_2/m_3>0$;.
In the cases (b) and (c), NH$\rightarrow$IH, the vertical axis is illustrated as
$1/\widetilde{x}$ instead of $\widetilde{x}$. 
Also, in the case (d), IH$\rightarrow$IH, the horizontal and vertical 
axes are illustrated by $1/x$ and $1/\widetilde{x}$, respectively.
\end{quotation}

\end{center}
\end{figure}


\vspace{4mm}
\begin{thebibliography}{99}
%
%
\bibitem{no-go}  Y.~Koide, {Phys. Rev. D}  {\bf 71} 
(2005) 016010.  For a review, Y.~Koide, arXiv:0801.3491 [hep-ph], 
Invited talk presented at International Workshop on Grand 
Unified Theories: Current Status and Future Prospects, 
Ritsumeikan University, Kusatsu, Shiga, Japan,
Dec. 17 - 19, 2007, to be published in the proceedings
(AIP Conf. Proc.).
%
%
\bibitem{Froggatt} C.~Froggatt and H.~B.~Nielsen, Nucl.~Phys. {\bf B147}
  (1979) 277.
%
\bibitem{23sym}
T.~Fukuyama and H.~Nishiura,
hep-ph/9702253, in Proceedings of {\it the International 
Workshop on Masses and 
Mixings of Quarks and Leptons},  Shizuoka, Japan, 1997, 
edited by Y.~Koide (World Scientific, Singapore, 1998), p. 252; 
R.~N.~Mohapatra and S.~Nussinov, Phys.~Rev. {\bf D60}, 013002,
(1999);
E.~Ma and  M.~Raidal, Phys.~Rev.~Lett. {\bf 87}, 011802 (2001); 
C.~S.~Lam, Phys.~Lett. {\bf B507}, 214 (2001); 
K.~R.~S.~Balaji, W.~Grimus and T.~Schwetz, 
Phys.~Lett. {\bf B508}, 301 (2001); 
W.~Grimus and L.~Lavoura, Acta Phys.~Pol. {\bf B32}, 3719 (2001).
%
%
%
%
%
%
\bibitem{MINOS} D.~G.~Michael {\it et al.}, MINOS collaboration,
Phys.~Rev.~Lett. {\bf 97} (2006) 191801;
J.~Hosaka, {\it et al.}, Super-Kamiokande collaboration, Phys.~Rev. 
{\bf D74} (2006) 032002.
%
\bibitem{KamLAND} S.~Abe, {\it et al.}, KamLAND collaboration,
arXiv:0801.4589.
%
\end{thebibliography}
\end{document}